\documentclass[preprint,12pt,authoryear]{elsarticle}

\usepackage{amssymb}
\usepackage{amsmath}
\usepackage{url}
\usepackage{multirow}
\usepackage{makecell}
\usepackage{booktabs}

\journal{Computer methods in applied mechanics and engineering}

\begin{document}

\begin{frontmatter}



\title{Image-based modelling of rock non-linear deformation under low-stress levels} %% Article title


\author[label1]{Rui Li} %% Author name
\author[label1]{Yi Yang}
\author[label1]{Wenbo Zhan}
\author[label2]{Jianhui Yang}
\author[label1,label3]{Yingfang Zhou\corref{cor1}}\ead{yingfang.zhou@abdn.ac.uk}

\cortext[cor1]{Corresponding author}
\affiliation[label1]{organization={School of Engineering, University of Aberdeen},
            postcode={AB24 3FX},
            country={UK}}
            
\affiliation[label2]{organization={Geoscience Research Centre, Total E\&P UK Limited},
            country={UK}}

\affiliation[label3]{organization={School of Energy Resources, China University of Geosciences},
            city={BeiJing},
            postcode={100083},
            country={China}}

%% Abstract
\begin{abstract}
Rock geophysical properties are widely reported to exhibit non-linear behaviours under low-stress conditions (below 10-20 MPa) before transitioning to the linear elastic stage, primarily due to the closure of microcracks and grain interfaces.
Image-based modelling of rock deformation struggles to effectively characterise the microcrack closure effect because of the partial-volume effect, where image voxels are larger than microcracks and contain both pore and solid phases.
This study presents a novel method to simulate non-linear rock deformation under elevated stress conditions. 
The method reconstructs digital rock models by treating partial-volume voxels as transitional phases that incorporate microcracks. 
By assigning intermediate elastic moduli and assuming that the pore portion within each partial-volume voxel deforms before the remaining solid content, the method employs the finite element method to simulate rock deformation and calculate the porosity of the deformed model.
The method is tested on two Bentheimer sandstone models, and the results demonstrate its ability to predict the non-linear changes in porosity and elastic properties as the effective stress increases. 
This work provides a new pathway for image-based modelling of non-linear rock deformation considering the microcrack closure effect, offering valuable insights into the complex mechanical behaviour of rocks under confinement.
\end{abstract}


\begin{highlights}
\item Rocks experience non-linear deformation at low-stress levels (below 10-20 MPa) due to the closure of microcracks.
\item A novel image-based modelling method is proposed to characterise the rock non-linear deformation.
\item This method is tested on sandstone models and successfully predicts the non-linear variations in rock porosity and elastic properties.
\end{highlights}

%% Keywords

\end{frontmatter}

\section{Introduction}
Studying the mechanical deformation processes of rocks is crucial for many engineering applications performed under low-stress levels (typically less than 10-20 MPa), such as oil and gas production in shallow formations \citep{Rezaei2020, Wang2021b, Aydin2021}, geothermal energy extraction \citep{Baria1999, Liu2022, Hu2024}, underground carbon dioxide storage \citep{Rutqvist2002, Bao2021}, and civil engineering design \citep{Cong2020, Bao2022, Asemi2024}. 
Many studies analysing stress-strain relationships from rock compression tests have shown that rocks, particularly sandstones, often undergo non-linear deformation before reaching a linear stage \citep{Lionco2000, Zhu2002, Corkum2007, Wang2021b, Lin2024}.
Additionally, the geophysical properties of rocks can change significantly during this non-linear deformation process. 
For example, properties such as porosity and permeability may drastically decrease \citep{Jasinski2015, AlBalushi2022a, Lin2024}, while elastic moduli and sound wave velocities tend to increase significantly \citep{Johnson1996, Saxena2017a, Hao2022}. 
The different behaviours of rock geophysical properties during non-linear deformation are primarily attributed to the closure of microcracks and grain boundaries under elevated stresses \citep{Mavko1978, Saxena2017, Lin2024, Peng2024}. 
The closure of these pore structures reduces the number of available flow paths, leading to a decrease in permeability. 
Simultaneously, as rock grains become more compacted, the rock rapidly becomes stiffer. 
Therefore, investigating non-linear deformation is essential for predicting changes in rock geophysical properties and understanding their implications for specific engineering applications.

Some mathematical models have been developed to characterise the non-linear stress-strain relationship under low-stress levels. 
These include logarithmic models \citep{Jones1975}, polynomial models \citep{Gangi1978}, exponential models \citep{McKee1988b, Clarkson2012}, power-law models \citep{Dong2010a}, and stress-sensitive theory models \citep{Shapiro2003, Shapiro2005, Liu2009a, Pervukhina2008}. 
A systematic evaluation of these models can be found in the work of \citet{Zhao2019c}.
With the advancement of high-resolution imaging techniques, pore-scale numerical modelling using digital rock images has become a routine process for predicting rock geophysical properties and simulating the complex physical processes within rocks \citep{Madonna2012, Andra2013b, Andra2013a, Blunt2013a, Ikeda2020a, Li2024}. 
This technique reconstructs models based on digital rock images and conducts numerical simulations on these digital models to predict the rock's geophysical properties.
Benefiting from high-resolution imaging techniques, this approach effectively captures intricate rock structures and serves as a non-destructive method for analysing rock properties. 
It provides a valuable alternative or supplement to conventional laboratory experiments and mathematical models.
Some studies have used image-based modelling methods to investigate non-linear rock deformation and estimate stress-dependent rock properties. 
For instance, \citet{Saenger2016} proposed a workflow to predict the elastic properties of sandstone under varying stresses. 
They extracted an intermediate phase between pore and solid and assigned varying elastic moduli (0\% to 100\% of the quartz mineral) to this phase, representing the rock's elastic moduli under different stresses. 
Furthermore, \citet{Dautriat2009} introduced a pore-scale compaction law to make the pores and throats stress-dependent within the pore network modelling method. 
This stress-dependent pore network modelling approach can estimate changes in porosity and permeability under confining stresses. 
Additionally, \citet{AlBalushi2022a} proposed a workflow that integrates the finite element method with the Lattice-Boltzmann method to model stress-dependent porosity and permeability in synthetic digital rock models.

Image-based modelling of rock deformation under low-stress levels presents significant challenges due to limitations in image resolution. 
During the model reconstruction process, a balance must be struck between image resolution and the field of view \citep{Schluter2014, Bazaikin2017, Jiang2024}. 
This trade-off arises from the constraints of imaging instruments and the considerable processing and computational costs associated with subsequent numerical simulations.
One major issue caused by this trade-off is the partial-volume effect, which occurs due to coarse image resolution \citep{Schluter2014, Gerke2015, Korneev2018, Li2024}. 
The partial-volume effect means that many image pixels are composed of multiple material phases rather than a single, pristine phase. 
In conventional digital rock physics workflows, models are typically reconstructed by treating each voxel as if it contains a single material phase.
Additionally, microcracks can be significantly smaller than the pixel size, making it difficult for digital rock images to properly capture these features \citep{Schluter2014, Andra2013b, Saxena2017a}.
In fact, numerous studies have reported that the modelled elastic properties of rocks are often significantly overestimated because many microcracks and micropores are not adequately resolved in the reconstructed models \citep{Madonna2012, Andra2013b, Saxena2017, FarhanaFaisal2019a, Goldfarb2022}.
Thus, it is crucial to incorporate the partial-volume effect into the model reconstruction process to ensure that models exhibit correct elastic properties at the initial state, prior to simulating rock deformation under stress.

Efforts have been made to mitigate the influence of the partial-volume effect during the model reconstruction process. 
One intuitive approach is to increase the image resolution to capture finer rock structures. 
Since using very high-resolution imaging instruments can be extremely expensive and time-consuming, some computational methods have been proposed to enhance the resolution of images scanned at relatively lower resolutions.
Techniques such as deep learning for generating super-resolution images have been proposed and achieved great success \citep{Wang2020c, Wang2021c, Niu2022, Buono2023}.
However, these super-resolution methods generally require rock images scanned at different resolutions to train deep learning models properly \citep{Wang2020c, Wang2021c, Alqahtani2022, Niu2022}. 
Furthermore, generating super-resolution models results in finer resolutions and larger model sizes, which can significantly increase simulation costs.
Another promising solution for addressing the partial-volume effect is to treat the partial-volume phase as a transitional phase between pristine phases \citep{Ikeda2020a, Goldfarb2022}. 
By determining the pore and solid volume fractions within these transition phases, mathematical models such as effective medium theory can be employed to assign the elastic moduli to these transition phases. 
Subsequently, the effective elastic properties of the entire rock model can be calculated. 
The key aspect of this approach is determining the local pore fraction for each partial-volume voxel by establishing a correlation between the pixel greyscale value and the pore fraction.
\citet{Goldfarb2022} and \citet{Ikeda2020a} established linear and power-law correlations between pixel greyscale values and pore fractions. 
They tested their methods on several Berea sandstone models and found that the modelled effective elastic moduli closely aligned with experimental measurements.

In this work, we propose a novel method to model rock deformation under low-stress levels. 
This method first extracts the partial-volume pixels and determines the pore fraction within each pixel using the cumulative Beta distribution function method. 
It uses the finite element method to simulate model deformation under fixed small strain conditions, assuming that the partial-volume pixels contain microcracks that close before the remaining solid phase. 
After each deformation simulation, the porosity of the deformed model is computed, and the deformed grid is regularised for a new deformation simulation routine.
We test this method on two Bentheimer sandstone models. 
The results demonstrate that our method successfully predicts the non-linear stress-strain relation under low-stress ranges before reaching the linear stage. 
Furthermore, the modelled stress-dependent elastic properties align well with laboratory measurements. 
This work provides a new pathway for image-based modelling of non-linear rock deformation under low-stress conditions. 
Future research can extend this method to model stress-sensitive transport properties and study pore size distribution under elevated stresses.

\section{Methodology}
\subsection{Digital rock models}\label{sec: digital rock images}
\begin{figure}
\centering
\includegraphics[width=\textwidth]{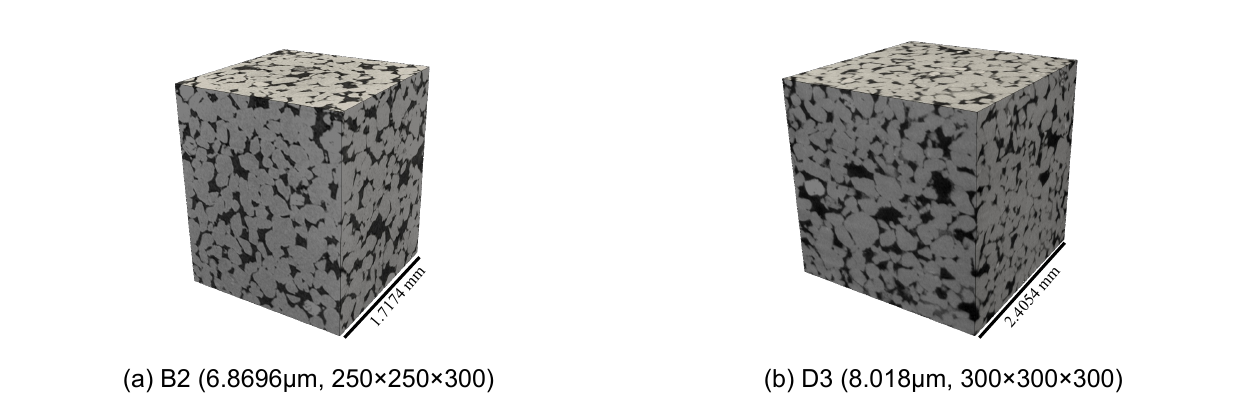}
\caption{Two Bentheimer sandstone digital models used in this work. 
B2 has a pixel size of $6.8696\mu m$, with a total volume size of $250\times 250\times 300$ pixels.
D3 has a pixel size of $8.018\mu m$, with a total volume size of $300\times 300\times 300$ pixels.
}
\label{Models}
\end{figure}

This work utilises rock images and laboratory measurement data obtained from \citet{liang_2020_3886416} (\url{https://doi.org/10.5281/zenodo.3886416}). 
According to \citet{Liang2020a}, the solid matrix of the two rock models is primarily composed of quartz, feldspar, and clay. 
The elastic properties and wave velocities of three Bentheimer sandstone samples were measured by \citet{Liang2020a}. 
The average porosity of these tested samples is 23.37\%, and the average density is 1980 $kg/cm^3$.
Digital rock images of core samples from the same rock blocks were also obtained. 
An analysis of the digital images of two rock samples (B2 and D3) shows that B2 contains 0.6\% clay and 1.69\% feldspar, while D3 has 2.16\% clay and 2.19\% feldspar in terms of volume fractions \citep{Liang2020a}. 
The digital rock images of B2 have a resolution of 3.4348 $\mu m$, with a size of $988\times 1012 \times 992$ pixels. 
In contrast, the digital images of the D3 sample have a resolution of 4.009 $\mu m$ and a size of $988\times 1012 \times 1000$ pixels.
In this work, we crop a $500\times 500\times 600$ sub-volume from the original B2 image stack, which is the maximum coverage of the rock images. 
We then resample the model to construct a $250\times 250\times 300$ model with a doubled pixel size of 6.8696 $\mu m$. 
Similarly, we crop a $600\times 600\times 600$ sub-volume from the original D3 model and double the resolution to form a $300\times 300\times 300$ model. 
The two constructed rock models can be seen in Figure \ref{Models}.

%-----------------------------------------------------------
%-----------------------------------------------------------

\subsection{Rock model reconstruction}
\label{Beta}
\subsubsection{Beta distribution method}
To model the deformation of rock models, it is crucial to reconstruct a digital rock model with appropriate elastic properties under zero confining pressures. 
Traditional methods for reconstructing rock models often result in significantly overestimated rock elasticity due to the inadequate treatment of partial-volume effects \citep{Madonna2012, Andra2013b, Saxena2017, FarhanaFaisal2019a}. 
Therefore, we employ a new technique for digital rock reconstruction that accounts for the partial-volume effect.
This method utilises the cumulative distribution function of the Beta distribution to represent the relationship between the solid fraction and the cumulative probability of voxels within rock models. 
The relationship is expressed as follows:
\begin{equation}
	F(x \vert \alpha, \beta) = \frac{1}{B(\alpha,\beta)} \int_0^x t^{\alpha-1} (1-t)^{\beta-1} dt, \quad x = [0,1] 
	\label{cdf_x}
\end{equation}
where $F(x\vert \alpha, \beta)$ represents the solid fraction, while $x$ denotes the cumulative probability of the voxels. 
$\alpha$ and $\beta$ are two positive hyperparameters that control the shape of the function.
Additionally, $B(\alpha,\beta)$ is the Beta function, which ensures that the total probability sums to 1.

To solve Equation \ref{cdf_x}, several constraints must be established. 
The first constraint ensures that the cumulative sum of Equation \ref{cdf_x} over the entire range [0,1], denoted as $CF$, conserves the total solid fraction and total porosity within the model. 
This relationship is expressed as:
\begin{equation}
	CF = \int_0^1 F(x \vert \alpha, \beta) dx = \int_0^1 \left[\frac{1}{B(\alpha,\beta)} \int_0^x t^{\alpha-1} (1-t)^{\beta-1} dt \right]dx = f_s = 1-\phi
	\label{total cdf}
\end{equation}
where $f_s$ is the total solid fraction and $\phi$ denotes the total porosity of the model.

Secondly, reference points are selected by incorporating Gaussian distributions for the solid and pore phases. 
The pure phases in digital rock images tend to exhibit a normal distribution, with the partial-volume phase falling within the range of the two mean values of these Gaussian distributions \citep{Goldfarb2022}.
Consequently, two constraints can be defined as follows:
\begin{eqnarray}
	F(x = P_1 \vert \alpha, \beta) = \frac{1}{B(\alpha,\beta)} \int_0^{P_1} t^{\alpha-1} (1-t)^{\beta-1} dt < f_{s, P_1} = \frac{1}{N(P_1)} \nonumber \\ 
	F(x = P_2 \vert \alpha, \beta) = \frac{1}{B(\alpha,\beta)} \int_0^{P_2} t^{\alpha-1} (1-t)^{\beta-1} dt > f_{s,P_2} = 1-\frac{1}{N(P_2)}
	\label{constraints}
\end{eqnarray}
where $f_{s,P_1}$ and $f_{s,P_2}$ are two solid fraction criteria that constrain the values of $F(x = P_1 \vert \alpha, \beta)$ and $F(x = P_2 \vert \alpha, \beta)$.
The terms $P_1$ and $P_2$ denote the mean intensity values of the Gaussian distributions for the pore and solid phases. 
Additionally, $N(P_1)$ and $N(P_2)$ represent the total number of voxels with intensities corresponding to the cumulative voxel fractions $P_1$ and $P_2$. 

Furthermore, the solution of Equation \ref{cdf_x} is approached numerically by defining a wide range for both $\alpha$ and $\beta$ and allowing the two parameters to iteratively change within that range. 
For each pair of $\alpha$ and $\beta$, we check if they satisfy Equations \ref{total cdf} and \ref{constraints}. 
However, it is possible that the two constraints in Equation \ref{constraints} may not be satisfied simultaneously. 
In some cases, one condition may be met while the other is difficult to satisfy.

When this occurs, the iteration process may continue, potentially resulting in excessively large values for $\alpha$ and $\beta$. 
This can lead to a wide region above $P_1$ becoming classified as a pure pore phase or a wide region below $P_2$ being categorised as a pure solid phase. 
To address this issue, we introduce a breakpoint in the workflow to halt the iteration process when the following condition is met:
\begin{equation}
	 \left\vert \frac{f_{s,P_2}^T-f_{s,P_2}}{f_{s,P_1}^T-f_{s,P_1}} \right\vert =
	\begin{cases}
		\geq \frac{N(P_1)}{N(P_2)} ,\quad f_{s,P_2}^T \geq f_{s,P_2} \ \& \ f_{s,P_1}^T > f_{s,P_1}\\
		\\
		\leq \frac{N(P_1)}{N(P_2)} ,\quad f_{s,P_1}^T \leq f_{s,P_1} \ \& \ f_{s,P_2}^T < f_{s,P_2} \\
	\end{cases}
	\label{breakpoint}
\end{equation}
where $f_{s,P_1}^T$ and $f_{s,P_2}^T$ represent the solid fractions at $P_1$ and $P_2$ generated by the current $\alpha$-$\beta$ pair during the iteration process.
This condition ensures that the iteration does not infinitely increase $\alpha$ and $\beta$ and produce unrealistic outcomes.

After obtaining the solid fraction distribution of the rock model, we separate the model into a predefined number of sub-phases and calculate the pore fractions for each sub-phase. 
During this process, we evenly partition the voxels between $P_1$ and $P_2$ based on their intensity levels. 
Voxels above $P_2$ are treated as pure quartz and feldspar, while those below $P_1$ are considered pure pore.
Next, we determine the cumulative volume fraction within each sub-phase, denoted as $CF_i$, using the following equation:
\begin{equation}
	CF_i = \int_{T_{i-1}}^{T_{i}} F_i(x=[T_{i-1},T_{i}] \ \vert \ \alpha, \beta) dx = \int_{T_{i-1}}^{T_{i}} \left[\frac{1}{B(\alpha,\beta)} \int_0^x t^{\alpha-1} (1-t)^{\beta-1} dt \right]dx
	\label{CF_i}
\end{equation}
where $i$ ranges from 1 to $N$, and we set the threshold $T_0$ to $P_1$.
The pore fraction of each sub-phase, $\phi_i$, can then be computed as:
\begin{equation}
	\phi_i =1- \frac{CF_i}{T_i-T_{i-1}}
	\label{porosity of subphase}
\end{equation}

\subsubsection{Elastic moduli allocation for sub-phases}
Once the pore fraction and volume fraction of each sub-phase are determined, we adopt an effective medium theory to compute their bulk and shear moduli. 
For simplicity, the minor occurrence of clay minerals in both models is neglected.
%Before doing so, it is essential to consider the clay phase, even though its total volume fraction is relatively small. 
%The minor occurrence of clay can significantly reduce the effective elastic properties of the rock \citep{Han1986, Liang2020a}.
%In this work, the clay phase is treated as being contained within the partial-volume phase, characterised by intensities between pore and pure quartz. 
%However, the precise distribution of clay minerals within this partial-volume phase is highly uncertain. 
%For simplicity, we assume that the clay content is homogeneously mixed with the quartz phase in the partial-volume phase. 
%Consequently, each sub-phase contains pore, clay, and quartz, with identical clay and quartz fractions across all partial-volume sub-phases.
Based on this assumption, we utilise the modified Hashin-Shtrikman upper and lower bounds to calculate the bulk and shear moduli of the partial-volume phases. 
The modified Hashin-Shtrikman bounds are expressed as follows:
\begin{eqnarray}
	&K_i^{\pm}&= K_1 + \frac{\phi_{i}/\phi_c}{(K_2-K_1)^{-1}+\frac{1-\phi_{i}/\phi_c}{K_1+4/3G_1}} \nonumber \\
	&G_i^{\pm}&= G_1 + \frac{\phi_{i}/\phi_c}{(G_2-G_1)^{-1}+\frac{2(1-\phi_{i}/\phi_c)(K_1+2G_1)}{5G_s(K_1+4/3G_1)}}
	\label{MHS1}
\end{eqnarray}
where the superscript $+/-$ denotes the upper and lower bounds, respectively; 
$\phi_c$ is the critical porosity, set at 36\% \citep{Goldfarb2022}. 
The upper and lower bounds are calculated by interchanging the materials labeled as 1 and 2. 
The upper bound is determined by treating the solid phase as material 1, while the lower bound is derived when the pore phase is designated as material 1.
%The bulk and shear moduli of the solid phase are calculated using:
%\begin{equation}
%	M_S=  \frac{1}{\frac{f_{C}}{M_{C}}+\frac{f_{Q}}{M_{Q}}}
%	\label{M_s}	
%\end{equation}
%where $M_S$ is the bulk or shear modulus of the solid matrix within the partial-volume sub-phases.
%$f_C$ and $f_Q$ are the fractions of the clay and quartz phases, and $f_C + f_Q = 1$.
%$M_C$ and $M_Q$ denote the bulk and shear moduli of pure clay and quartz, with their values provided in Table \ref{Parameters}.
The bulk and shear moduli of different phases can be seen in Table \ref{Parameters}.

%-----------------------------------------------------------
\begin{table}
\centering
\caption{Elastic properties of different components\label{Parameters}}
\begin{tabular}{cccc}
\toprule
   & Bulk modulus (GPa) & Shear modulus (GPa)  & Reference\\ \midrule
\multicolumn{1}{l}{Void} & 0  & 0 & \\
\multicolumn{1}{l}{Clay} & 12 & 6  & \citet{Vanorio2003}\\
\multicolumn{1}{l}{Quartz} & 37 & 44 & \citet{Mavko2009}\\
\multicolumn{1}{l}{Feldspar} & 37.5 & 15 & \citet{Mavko2009}\\
\bottomrule 
\end{tabular}
\end{table}
%-----------------------------------------------------------

Equation \ref{MHS1} defines the upper and lower elasticity bounds of a mixture of two materials. 
Typically, the effective elastic properties of rocks fall within these two extreme bounds \citep{Andra2013a}. 
In the case of sandstone, macro pores represent the predominant pore space. 
Consequently, partial-volume voxels are often situated at the boundaries of these macro pores, forming contact phases between pure solid and pore phases \citep{Li2024}.
In these contact phases, grains interact with the pore phase through numerous asperities, resulting in an intermediate overall stiffness that lies between the stiffest and softest mixtures \citep{Goldfarb2022}. 
As a result, the spatial distribution of pores and solids within the partial-volume phase is relatively uniform. 
This leads to the effective elastic properties of the partial-volume phase being systematically lower than the upper bound.
In this work, we compute the bulk and shear modulus of the partial-volume phases in sandstone models, denoted as $M_i$, by averaging their upper and lower bounds:
\begin{equation}
	M_i = \frac{1}{2}(M_i^{+}+M_i^{-}) \label{MHS2}
\end{equation}

\subsection{Rock deformation simulation and elastic properties calculation}
\label{FEM}
We employ a standard static finite element method (FEM) developed by \citet{Garboczi1995a} to calculate the effective elastic properties of the rock and model its deformation. 
This numerical technique is well-suited for structured grid systems and can be directly applied to models generated from digital rock images by treating each voxel as a finite element.
By imposing appropriate boundary conditions, the FEM computes elastic displacements, stress, and strain within each finite element. 
This is achieved through solving a variational formulation of the linear elastic equation using an iterative algorithm, such as the conjugate gradient method \citep{Arns2002a}. 
In this work, a periodic boundary condition is applied to eliminate any boundary effects arising from additional stresses.
To compute the effective elastic properties of the rock, macro strains along six different directions are individually applied, generating various stress distributions. 
These six macro strains will yield different stiffness components, allowing us to compute the effective bulk and shear modulus of the entire model \citep{Madadi2009a}. 
Subsequently, the compressional and shear wave velocities of the model are calculated using the following formulas:
%-----------------------------------------------------------
\begin{equation}
	V_p=\sqrt{\frac{K+\frac{4}{3}G}{\rho}} \hspace{0.2in} and \hspace{0.2in} V_s=\sqrt{\frac{G}{\rho}}
\label{Equation Vp_Vs}
\end{equation}
%-----------------------------------------------------------
where $\rho$ is the density of the model, $V_p$ is the compressional wave velocity, $V_s$ denotes the shear wave velocity, and $K$ and $G$ are the effective bulk and shear moduli calculated by the FEM.

\begin{figure}
\centering
\includegraphics[width=.5\textwidth]{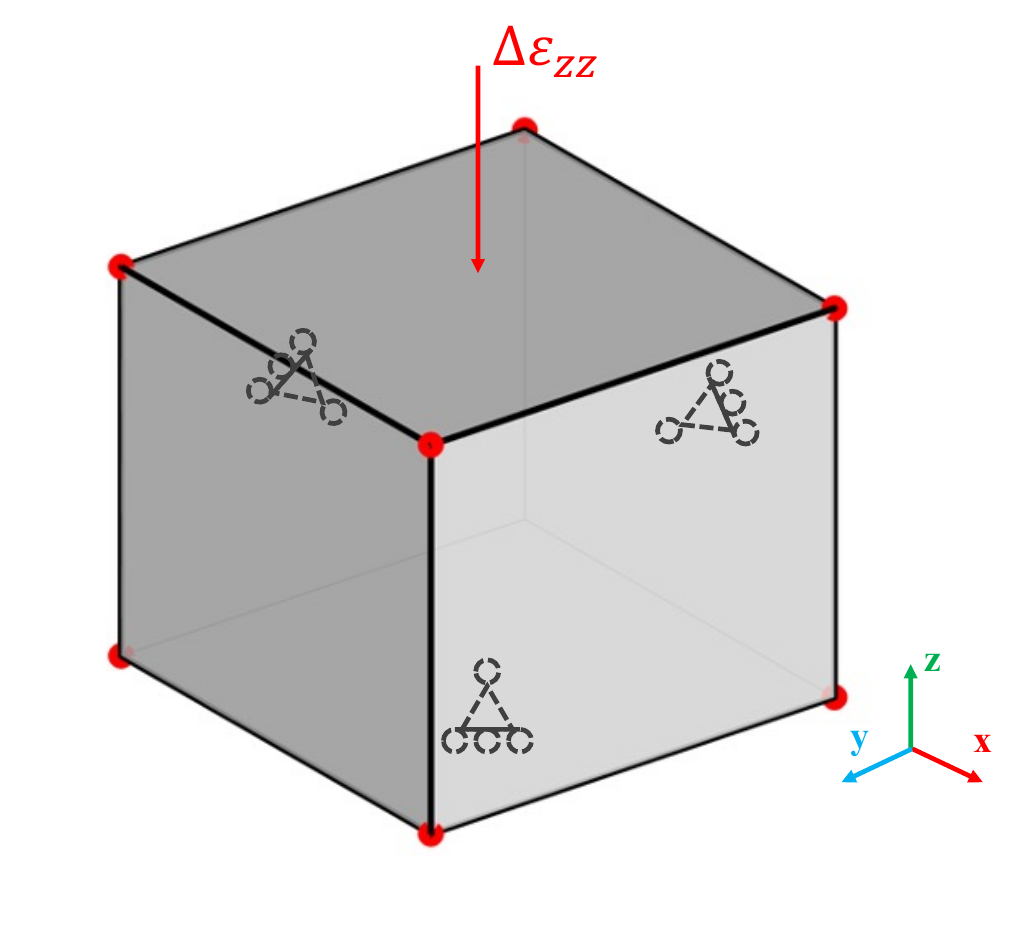}
\caption{To model rock uniaxial deformation, a macro strain is applied to the z+ face. 
The normal displacements of the x-, y-, and z- faces are fixed, while the x+ and y+ faces are allowed to deform freely. 
}
\label{FEM_strain}
\end{figure}

Additionally, we use the same FEM method to model rock deformation. 
However, in this case, a small uniaxial macro strain is applied to one face (the z+ face in this work) to simulate the uniaxial compression test typically conducted in laboratory. 
Other boundary conditions are set such that the normal displacements of x-, y-, and z- faces are fixed, while the x+ and y+ faces are allowed to deform freely. 
These boundary conditions are illustrated in Figure \ref{FEM_strain}.
A key hypothesis here is that the softer phase (pore space) within a mixed voxel deforms prior to the stiffer phase (solid phase) during the simulation of rock model deformation. 
This assumption is rational because partial-volume voxels in sandstone rocks are typically located around macro pore spaces, microcracks, and grain interfaces \citep{Li2024}. 
As a result, these voxels often partially cover pores rather than encompassing complete pores. 
These pore structures provide minimal resistance to external loads until they are fully closed, which is a primary cause of the non-linear deformation observed in rocks. 
Therefore, the partial-volume sub-phases can be treated as behaving like pure pore phases until their pore fractions diminish to zero. 
This assumption allows us to capture the non-linear deformation behaviour associated with the closure of microcrack and pore space in porous rocks.
%Similarly, the clay-quartz mixed voxels will act as pure clay until their deformed volume exceeds their clay fractions.
Based on this assumption, the FEM solves the variational formulation of the linear elastic equation to compute the displacements at each node. 
With the obtained node displacements and coordinates, we can then calculate the pore fraction of the partial-volume voxels and the total porosity in the deformed model, which will be discussed in the following sections.

%-----------------------------------------------------------
%-----------------------------------------------------------

\subsection{Pore fraction computation and model regularisation}
\label{Remap}
The FEM simulates local displacements at the nodes of each voxel within rock models under external stress and/or macro strain boundary conditions. 
After the deformation simulation, it is essential to calculate and update the pore fraction within each partial-volume voxel to create a new model. 
Additionally, to enable further deformation simulations using the lattice-based FEM method, the deformed grid system must be regularised. 
This new model will then be employed to simulate deformation under a new external macro strain boundary condition.

\subsubsection{Pore fraction computation}
\begin{figure}
\centering
\includegraphics[width=\textwidth]{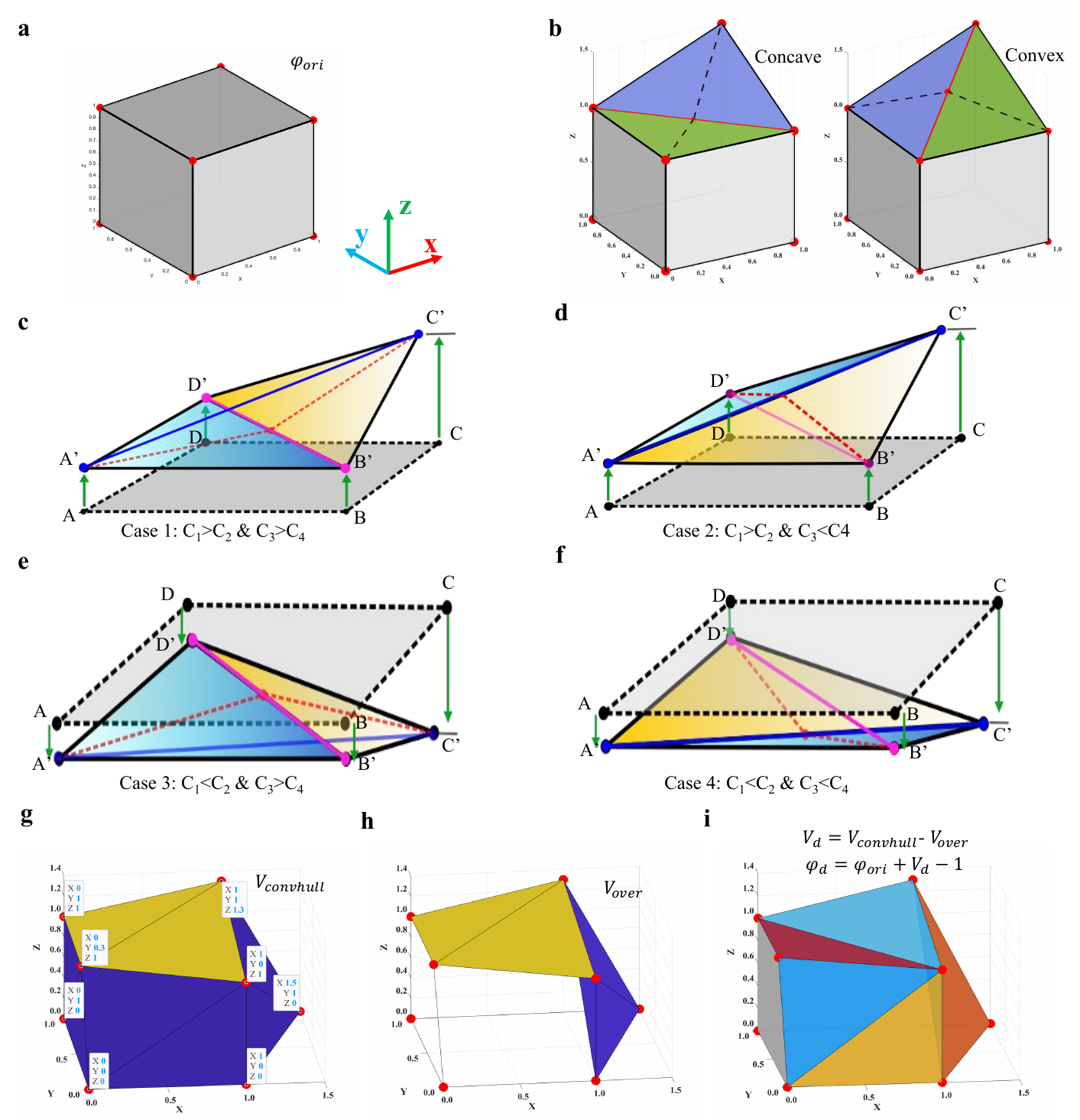}
\caption{The workflow for tracing changes in pore fractions within each voxel of digital rock models during deformation simulations.
(a) The original voxel with a pore fraction of $\phi_{ori}$,
(b) two possible cases of the shape composed of the four nodes in a cubic voxel after deformation,
(c-f) four cases that four nodes on a $z+$ surface can form,
(g) the 'convhull()' function in MATLAB calculates the maximum volume ($V_{convhull}$) that can be formed by the eight corner nodes of a voxel,
(h) all overestimated volumes are identified, and their total volume is denoted as $V_{over}$,
(i) the true volume of the voxel after deformation, $V_{d}$, is determined by subtracting $V_{over}$ from $V_{convhull}$, and the pore fraction is calculated using the equation $\phi_d=\phi_{ori}+V_{d}-1$
}
\label{Poro_Tracing}
\end{figure}

To calculate the pore fraction of partial-volume voxels after deformation, it is crucial to determine the volume of each voxel after the deformation. 
The displacements at the eight corner node points of the voxels are used to compute the deformed voxel volume.
Figure \ref{Poro_Tracing} illustrates the method for computing the deformed voxel volume and its associated pore fraction. 
Initially, each voxel has a volume of 1 and a pore fraction of $\phi_{ori}$ (see Figure \ref{Poro_Tracing}(a)).
After deformation, the coordinates of the eight corner nodes are obtained from the finite element method.
Using the 'convhull()' function in MATLAB, the volume of the deformed voxel can be calculated, as shown in Figure \ref{Poro_Tracing}(g).
The 'convhull()' function, based on a convex assumption, computes the maximum volume that can be formed by the eight nodes.
However, different assumptions regarding the volume enclosed by these nodes will affect the volume calculation, thereby influencing the pore fraction of the deformed voxel. 
The convex assumption tends to yield a larger volume than a concave one, potentially leading to an overestimation of the deformed voxel's volume. 
A typical example is illustrated in Figure \ref{Poro_Tracing}(b), where the volume defined by the four nodes on the z+ surface can represent either a concave or convex surface. 
The convex case results in a greater volume compared to the concave case.
To address this issue, we introduce a process to exclude the overestimated volume from the fully convex hull assumption. 
The deformed shape can have four scenarios. 
Taking the z+ surface ($S_{ABCD}$) as an example, the four scenarios are depicted in Figures \ref{Poro_Tracing}(c)-(f).
After deformation, the coordinates of the four node points ($A'(x_{A'}, y_{A'}, z_{A'})$, $B'(x_{B'}, y_{B'}, z_{B'})$, $C'(x_{C'}, y_{C'}, z_{C'})$, and $D'(x_{D'}, y_{D'}, z_{D'})$) are utilised to define four parameters that determine which case applies to the current surface. 
The four parameters are defined as follows:
\begin{eqnarray}
	C_1 &=& I_{A'}+I_{C'}\nonumber \\
	C_2 &=& I_{B'}+I_{D'}\nonumber\\
	C_3 &=& \lvert I_{A'}-I_{avg}\rvert + \lvert I_{C'}-I_{avg}\rvert \nonumber\\
	C_4 &=& \lvert I_{B'}-I_{avg}\rvert + \lvert I_{D'}-I_{avg}\rvert
\end{eqnarray}
where $I_{A'}$-$I_{D'}$ denote the coordinate ($I=x, y, or z$) of points $A'$-$D'$.
$I_{avg}$ is the average value of $I_{A'}$ to $I_{D'}$.
The parameters $C_1$ to $C_4$ are used to differentiate among the four cases, as shown in Figure \ref{Poro_Tracing}(c)-(f).
Table \ref{Concave faces} provides a summary of the conditions for the four cases that should be excluded from the convex hull volume $V_{convhull}$.

%-----------------------------------------------------------
\begin{table}
 \caption{Cases that need to be excluded from the 'convhull()' function in Matlab and their conditions}
 \centering
 \label{Concave faces}
 \begin{tabular}{ccc}
 \hline
   Case  &  Conditions & Surfaces need exclude\\
 \hline
 Case 1 & $C_1>C_2$ and $C_3>C_4$  & x+, y+, and z+ \\
 Case 2 & $C_1>C_2$ and $C_3<C_4$  & x-, y-, and z- \\
 Case 3 & $C_1<C_2$ and $C_3>C_4$  & x+, y+, and z+ \\
 Case 4 & $C_1<C_2$ and $C_3<C_4$  & x-, y-, and z- \\
 \hline
 \end{tabular}
 \end{table}
%----------------------------------------------------------- 

To calculate the pore fraction of a voxel after deformation, we first compute the volume of the convex hull formed by the eight corner points, denoted as $V_{convhull}$ (see Figure \ref{Poro_Tracing}(e)).
This volume represents the maximum possible volume that can be created by these nodes. 
Given that this convex hull volume can lead to overestimation, we need to identify specific cases where the convex assumption may not accurately reflect the true shape of the deformed voxel.
We achieve this by checking the conditions for all six surfaces of the voxel to determine which configurations result in overestimated volumes. 
The total of these identified overestimated volumes is represented as $V_{over}$.
After determining both the convex hull volume and the overestimated volume, we can compute the deformed volume $V_d$ and the updated pore fraction $\phi_{d}$ using the following equations:
\begin{eqnarray}
	V_d = V_{convhull}-V_{over} \nonumber \\
	\phi_d = \phi_{ori}+V_d-1
	\label{porosity calculation}
\end{eqnarray}

By calculating the pore fractions of all voxels in the model using Equation \ref{porosity calculation}, we can subsequently determine the total porosity of the deformed model.

%-----------------------------------------------------------
%-----------------------------------------------------------

\subsubsection{Model regularisation and refinement}
\begin{figure}
\centering
\includegraphics[width=\textwidth]{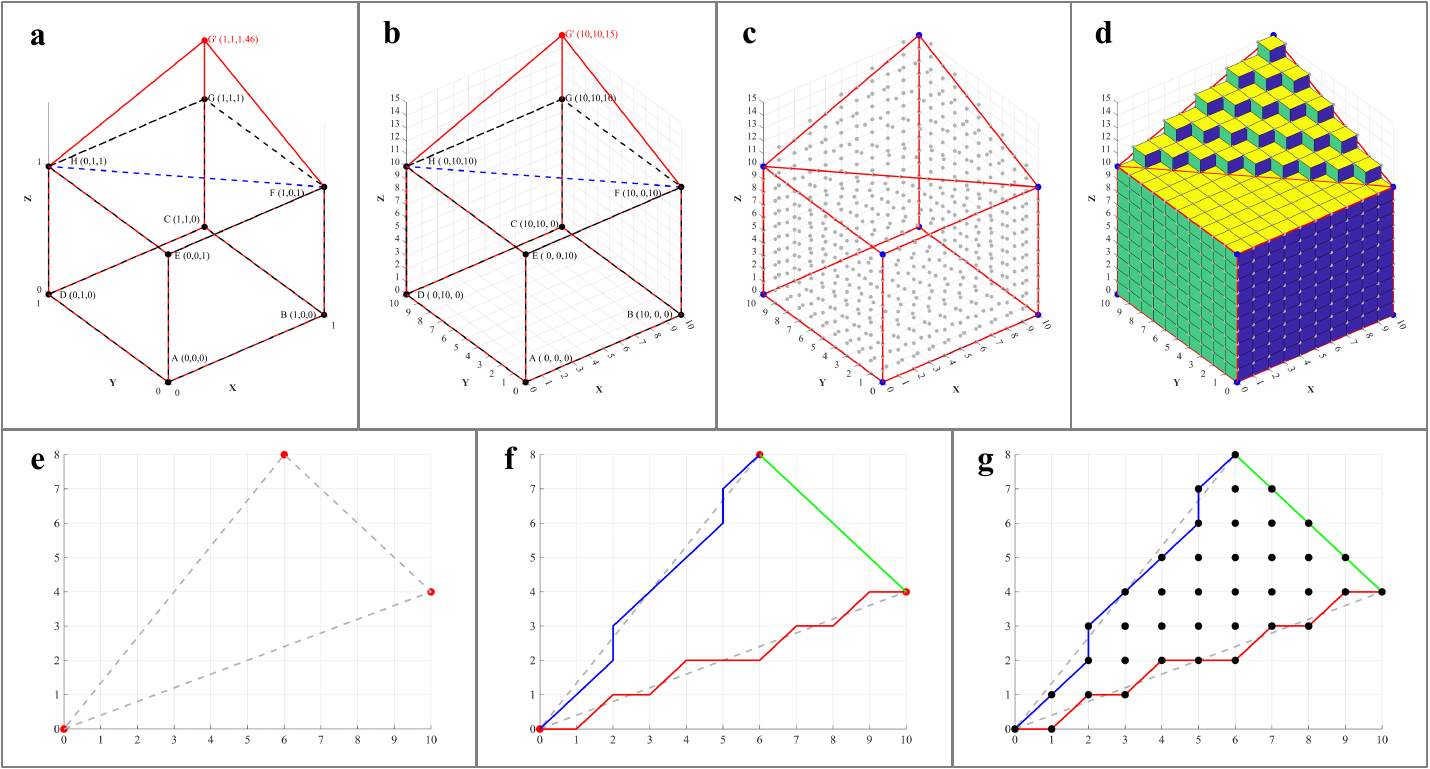}
\caption{The process of grid regularisation and refinement of the deformed model.
(a) the original grid, represented by the cube with dashed black edges and the corner point $G(1,1,1)$, alongside the   deformed grid, represented by the deformed cube with red edges and the displaced point $G'(1,1,1.46)$),
(b) the original grid system is refined onto a grid that is 10 times larger, and the coordinates of all nodes are rounded to align with the elevated grid system,
(c) all points on the boundary surfaces are identified,
(d) all cubes enclosed by the boundary points are identified,
(e) a triangle on a 2D grid system,
(f) the three edges of the triangle are converted to grid lines,
(g) all grid points enclosed by the edges are determined.}
\label{Regularisation}
\end{figure}

After deformation, the originally regular grids become irregular, necessitating a regularisation process before any further simulations can be conducted on the deformed model. 
This regularisation process is illustrated in Figure \ref{Regularisation}.
The coordinates of the voxel nodes change in the deformed model (as shown in Figure \ref{Regularisation}(a)). 
The first step is to allocate the deformed nodes onto an $n$-times finer grid system. 
A typical example is depicted in Figure \ref{Regularisation}(b), where the deformed voxel is placed on a grid that is 10 times finer. 
Additionally, the coordinates of the eight nodes are rounded to align with the nearest grid points. 
This rounding introduces some uncertainty into the grid regularisation.
However, we can anticipate that using a finer grid system can yield a more accurate regularised model.

Next, we determine the two spatial triangles on each voxel surface using the method outlined in Figure \ref{Poro_Tracing}. 
Once these triangular surfaces are defined, we can identify all points enclosed by the three edges of each triangle. 
Figures \ref{Regularisation}(e)-(f) illustrate this process. 
Initially, we convert the straight triangle edges (Figure \ref{Regularisation}(e)) into grid lines (Figure \ref{Regularisation}(f)). 
This allows us to locate all points within the grid edges (Figure \ref{Regularisation}(g)).
By extending this approach into three-dimensional space, we can identify all points on the surfaces of the deformed voxel (Figure \ref{Regularisation}(c)) and subsequently determine all cubes contained within these surfaces (Figure \ref{Regularisation}(d)). 
This step is repeated for all voxels within the model.
The irregular deformed model can therefore be remapped onto a regular grid system with a redefined resolution. 
In this work, we focus on studying deformation under low-stress levels.
Therefore, the extent of deformation is very small, and we do not alter the grid size during the regularisation process.

%-----------------------------------------------------------
%-----------------------------------------------------------

\subsection{Rock compression simulation workflow}
\begin{figure}
\centering
\includegraphics[width=0.5\textwidth]{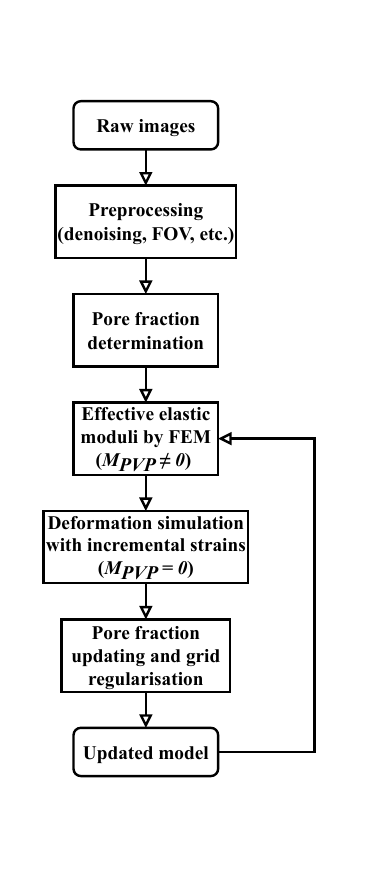}
\caption{The workflow of conducting Image-based modelling of rock deformation under low-stress levels.
FOV denotes the field of view, and $M_{PVP}$ represents bulk and shear moduli of the partial-volume phases.
}
\label{Workflow}
\end{figure}

The complete workflow for modelling rock deformation under low-stress levels is illustrated in Figure \ref{Workflow}. 
The process begins with the acquisition of raw rock images, followed by preprocessing steps, such as denoising and selecting the field of view. 
Next, the Beta distribution process is applied to determine sub-resolution pore fractions and segment the rock models into specific sub-phases (Section \ref{Beta}).
The finite element method is then employed to calculate the initial effective elastic properties of the rock model (Section \ref{FEM}). 
It is essential to note that the bulk and shear moduli of the partial-volume phases used in this step are computed directly from the effective medium theory models (Equations \ref{MHS1} and \ref{MHS2}). 
Following this, the FEM is utilised again to simulate the rock deformation under a predefined macro strain condition (Section \ref{FEM}). 
During this simulation, the bulk and shear moduli of the partial-volume phases are assigned to those of the pore phase.
After the deformation simulation, the pore fraction of each voxel in the deformed model is recalculated, and the deformed model is regularised (Section \ref{Remap}). 
Finally, with the updated model, we can again compute the rock's effective elastic properties and simulate further rock deformation.
It is crucial to note that the macro strain applied to the rock model during deformation simulations must be very small. 
Therefore, within each increment of macro strain, we can reasonably assume that the model deforms linearly. 
When these small increments of macro strain accumulate, the overall rock deformation becomes non-linear due to the continuous changes in phase volume fractions with each iteration.

%-----------------------------------------------------------
%-----------------------------------------------------------

\section{Results and Discussion} 
\subsection{Reconstructed rock models}
\begin{figure}
\centering
\includegraphics[width=\textwidth]{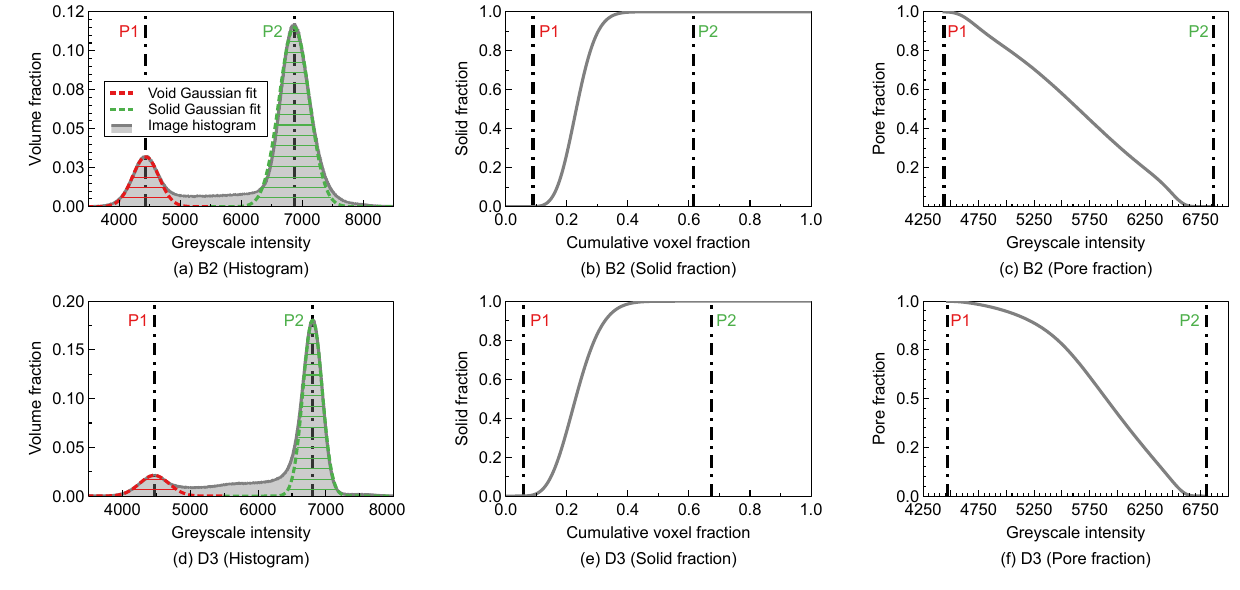}
\caption{The segmentation of rock models using the Beta distribution method.
(a, d) taking the mean values of pore and solid Gaussian distributions, denoted as $P_1$ and $P_2$, which are used to extract the partial-volume phase,
(b, e) the modelled solid fraction distribution within the partial-volume phase using Equations \ref{total cdf}-\ref{breakpoint},
(c, f) the relationship between sub-resolution pore fraction and greyscale intensity generated from the Beta distribution method.
}
\label{Beta seg}
\end{figure}

\begin{figure}
\centering
\includegraphics[width=\textwidth]{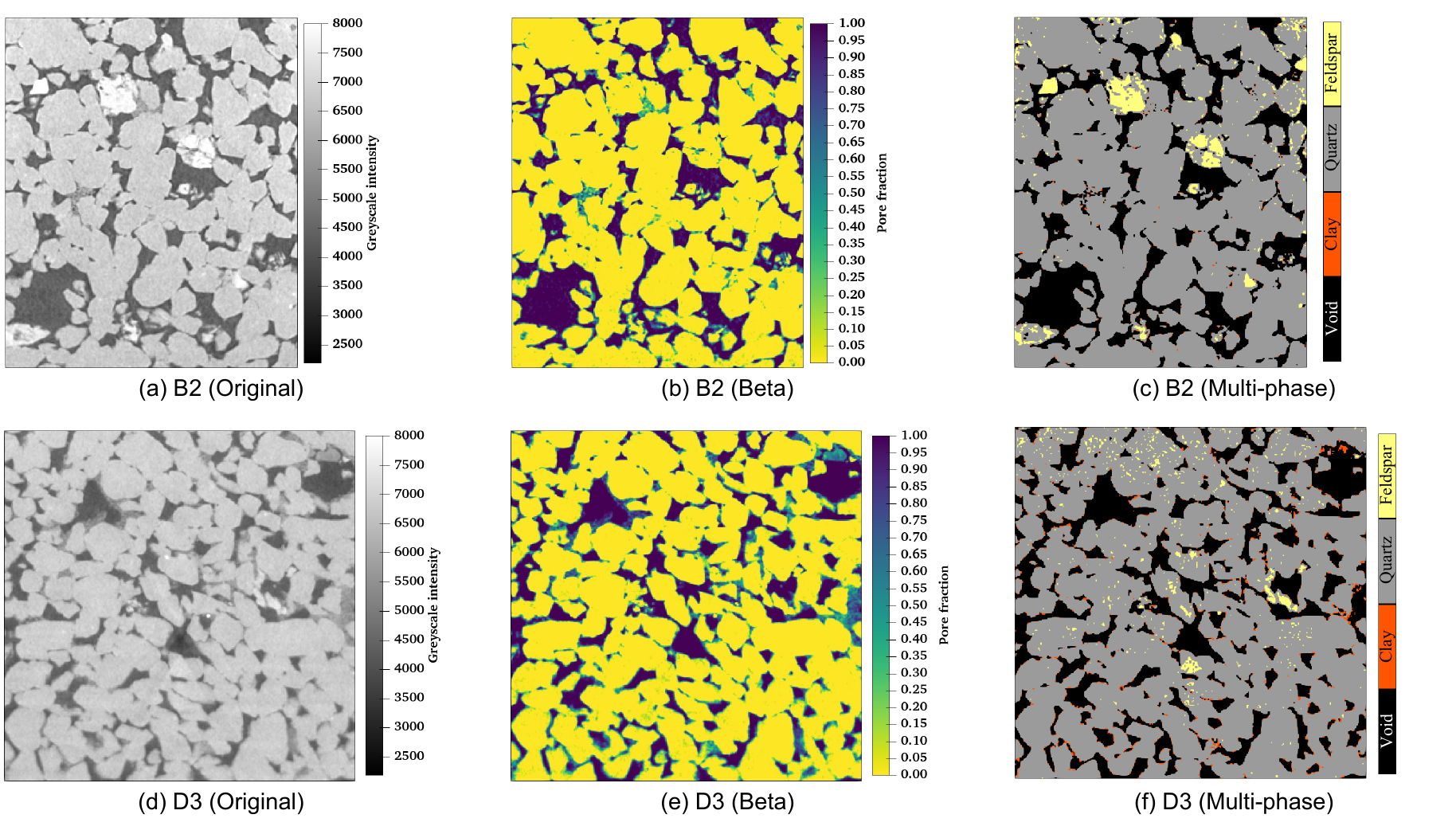}
\caption{The cross sections of models B2 and D3 before and after the segmentation process.
(a, d) the cross sections of the original models,
(b, e) the cross sections of the models (pore fraction distribution) after applying the Beta distribution segmentation method to divide the partial-volume phase into 500 sub-phases.
(c, f) the cross sections of the models after using a conventional thresholding-based method to segment the them into multi phases, including void/pore, clay, quartz, and feldspar.
}
\label{Segmented}
\end{figure}

Figure \ref{Beta seg} depicts the segmentation of rock models using the Beta distribution method.
The mean values ($P_1$ and $P_2$) of the pore and solid Gaussian distributions are selected to distinguish the partial-volume phase from the pure pore and solid phases (Figures \ref{Beta seg} (a) and (d)).
The solid fraction distribution is simulated using Equations \ref{total cdf}-\ref{breakpoint}.
The modelled solid fraction distributions for B2 and D3 plotted in Figures \ref{Beta seg} (b) and (e). 
Furthermore, Figures \ref{Beta seg} (c) and (f) show the relationship between the sub-resolution pore fraction and the greyscale intensity of partial-volume voxels.
Based on the solid fraction distribution, we divide the partial-volume phase of both B2 and D3 into 500 sub-phases. 
The pore fraction of each sub-phase is calculated using Equations \ref{CF_i} and \ref{porosity of subphase}. 
A thorough thresholding approach is also employed to segment the feldspar phase, in accordance with the feldspar volume fractions discussed in Section \ref{sec: digital rock images}.
Cross-sections of the original and segmented models are depicted in Figure \ref{Segmented}. 
The results show that the partial-volume voxels identified by the Beta distribution method are primarily located at the solid grain boundaries (Figures \ref{Segmented}(b) and (e)), creating a transition region between the solid grains and the pure pore phase. 
Importantly, the narrow grain interfaces are effectively represented by the partial-volume phase in the Beta distribution method, whereas many grain interfaces are lost in the conventional multi-phase segmentation method (Figures \ref{Segmented}(c) and (f)).

%-----------------------------------------------------------
%-----------------------------------------------------------

%-----------------------------------------------------------
%-----------------------------------------------------------

\subsection{Rock deformation under elevated stresses}
\begin{figure}
\centering
\includegraphics[width=\textwidth]{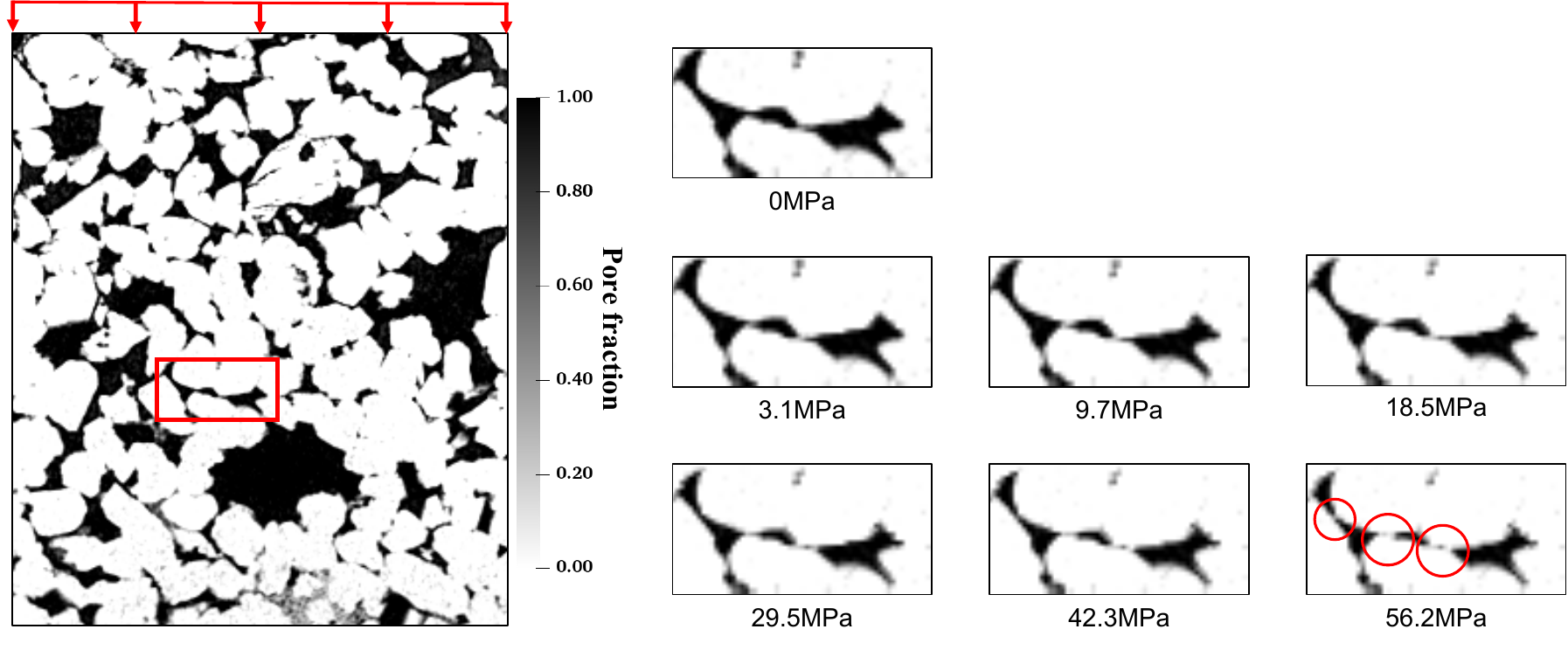}
\caption{2D views of rock deformation in model B2 under elevated stress conditions, with zoomed-in views that highlight the changes in rock structure as stress increases from 0 to 56.2MPa.
Arrows indicate the direction of the macro strain conditions applied during the finite element method simulations.
}
\label{Closure_2D}
\end{figure}

\begin{figure}
\centering
\includegraphics[width=\textwidth]{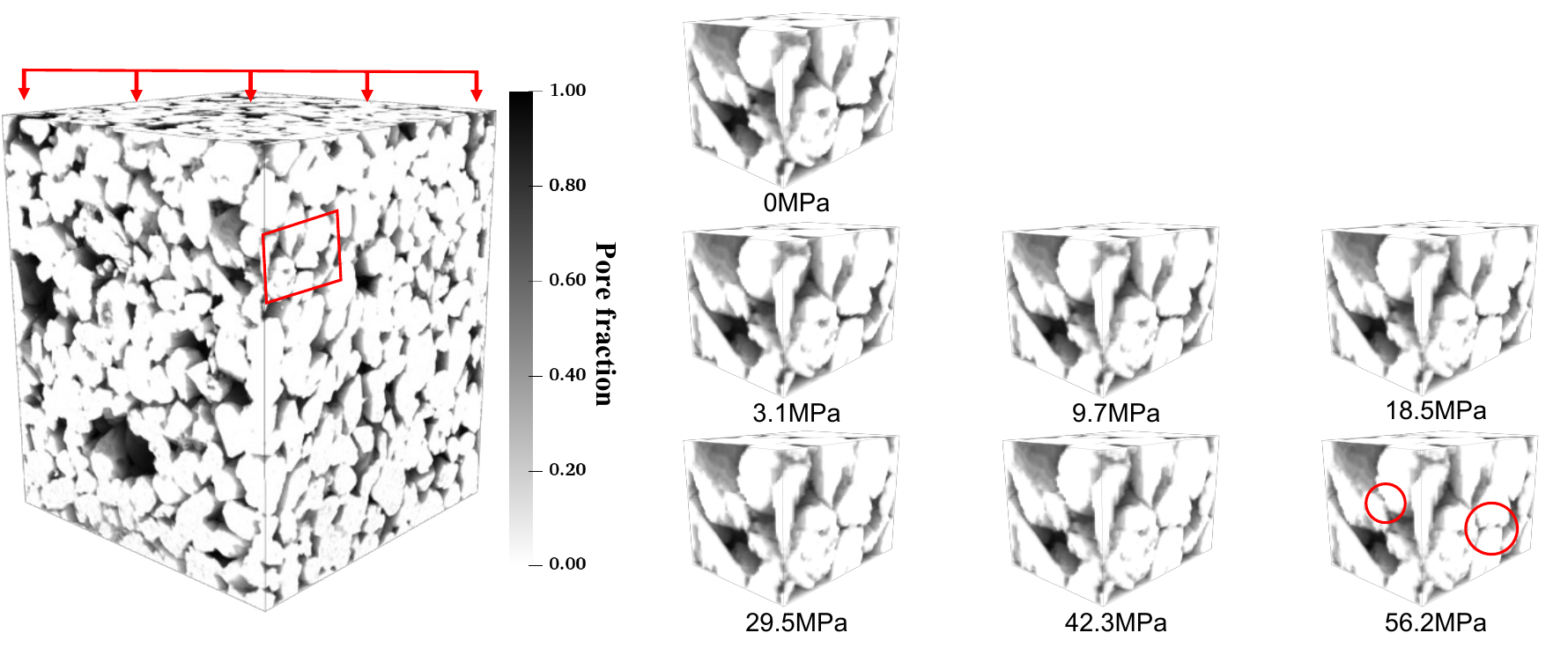}
\caption{3D views of rock deformation in model B2 under elevated stress conditions with zoomed-in sections highlighting the changes in the rock structure as stress increases from 0 to 56.2 MPa. 
Arrows indicate the direction of the macro strain conditions applied during the finite element method simulations.
}
\label{Closure_3D}
\end{figure}

Figures \ref{Closure_2D} and \ref{Closure_3D} illustrate the deformation of model B2 as uniaxial stress increases from 0 to 56.2 MPa. 
For a more dynamic visualisation of the model deformation under varying stress levels, animations are provided in the supplementary documents. 
From these figures, it is evident that the pore space diminishes as stress increases, leading to the closure of some narrow pore spaces. 
Consequently, connected pore spaces become separated into several smaller pore spaces. 
Therefore, originally well-connected pores become disconnected or exhibit reduced connectivity.
The observed reduction in pore connectivity provides insight into the significant decrease in permeability that occurs in rock samples as effective stress increases \citep{Fatt1952a, AlBalushi2022a, Dautriat2009}.

%-----------------------------------------------------------
%-----------------------------------------------------------

%-----------------------------------------------------------
%-----------------------------------------------------------

\begin{figure}
\centering
\includegraphics[width=\textwidth]{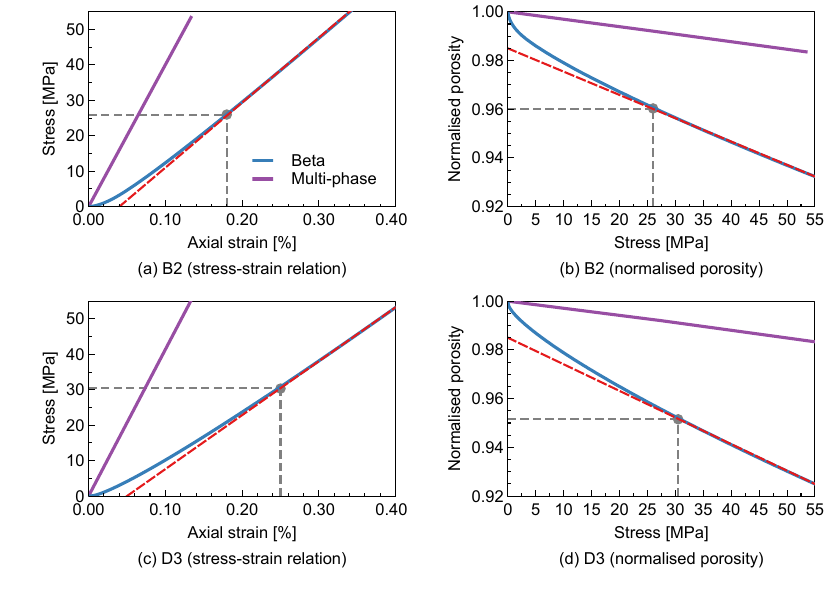}
\caption{Stress-strain relationships of models B2 (a) and D3 (c).
(b) and (d) are normalised porosity changes with elevated stresses of models B2 and D3.
Red dashed lines represent the linear sections in each subfigure.
Cross intersections of the two grey dashed lines represent the separation of the non-linear and linear deformation stages.
}
\label{Porosity}
\end{figure}

Figure \ref{Porosity} presents the stress-strain relationship and the normalised porosity (the porosity under different stress levels divided by the initial porosity at 0MPa) as stress increases. 
For both the stress-strain curve and the normalised porosity, the rock compression modelled using the Beta distribution segmentation shows a distinct non-linear behaviour before transitioning to a linear phase. 
In contrast, the rock compression using the conventional multi-phase model exhibits a consistently linear response.
This demonstrates that the proposed method more accurately captures the non-linear deformation that occurs at low-stress levels. 
During the compression simulations, the model segmented by the Beta distribution contains numerous micro cracks and micro pore spaces at grain interfaces, which separate the solid grains. 
As a result, a small amount of stress is sufficient to induce the initial deformation under a given macro strain. 
However, as these micro pore spaces close, the solid grains make more direct contact, and the solid framework becomes more compacted. 
Consequently, the same macro strain requires progressively higher stress. 
This gradual stiffening effect explains the presence of the non-linear section in both the stress-strain relationship and the normalised porosity.
Once the micro-pores are mostly closed, the rock model stabilises, and further changes in porosity primarily occur within the larger, macro-scale pores. 
At this stage, both the stress-strain relationship and the normalised porosity follow a linear trend.
The non-linear deformation observed in the Beta-distribution-based model cannot be properly captured in conventional multi-phase models. 
Moreover, the model generated by the conventional method is significantly stiffer compared to the one produced using the proposed method.

%-----------------------------------------------------------
%-----------------------------------------------------------

\begin{figure}
\centering
\includegraphics[width=\textwidth]{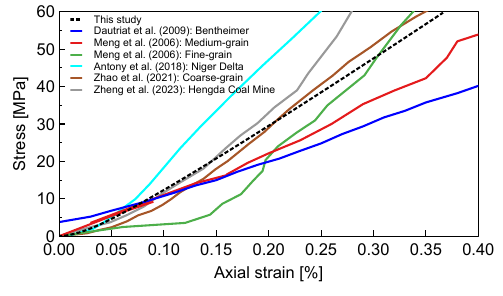}
\caption{Comparison of stress-strain curves between the simulation and documented experimental measurements.
The experimental results include: Bentheimer sandstone \citep{Dautriat2009}, medium- and fine-grain sandstone \citep{Meng2006}, Niger Delta sandstone \citep{Antony2018}, coarse-grain sandstone \citep{Zhao2021}, and sandstone retrieved from Hengda Coal Mine in China \citep{Zheng2023}.
}
\label{Stress-strain validation}
\end{figure}

Figure \ref{Stress-strain validation} presents a comparison between the simulated stress-strain curve and documented uniaxial test results from various sandstone samples. 
The experimental data includes measurements from: Bentheimer sandstone \citep{Dautriat2009}, medium- and fine-grain sandstone \citep{Meng2006}, Niger Delta sandstone \citep{Antony2018}, coarse-grain sandstone \citep{Zhao2021}, and sandstone retrieved from Hengda Coal Mine in China \citep{Zheng2023}.
All sandstone samples exhibit characteristic non-linear deformation behaviour, though the degree of non-linearity varies significantly depending on rock type, grain size distribution, and microcrack density. 
Despite these variations across different rock samples, the proposed methodology successfully captures the essential features of the non-linear deformation stage in the stress-strain response.

%-----------------------------------------------------------
%-----------------------------------------------------------

\begin{figure}
\centering
\includegraphics[width=\textwidth]{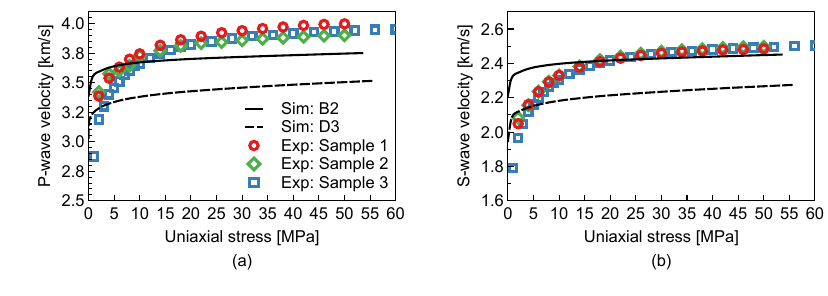}
\caption{P- (a) and S-wave (b) velocities of rock models under elevated stresses.
Scattered points represent experimental measurements taken from \citet{liang_2020_3886416}.}
\label{Velocity}
\end{figure}

Figure \ref{Velocity} depicts the P- and S-wave velocities of the rock models under increasing stress. 
Comparing the simulation results to experimental measurements reveals that the simulation effectively captures the significant increase in rock stiffness at low-stress levels. 
Overall, the modelled results align well with experimental data, though a gap remains between the two datasets. 
The P- and S-wave velocities of model D3 are lower than those of model B2, reflecting the higher clay and feldspar content in D3. 
This suggests that even small variations in phase volume fractions can significantly impact the modelled results. 
Thus, the representativeness of the rock models is a key uncertainty affecting simulation accuracy. 
%However, given the homogeneous nature of the sandstone samples, the discrepancy between the simulation and experimental results is not substantial.

\begin{figure}
\centering
\includegraphics[width=\textwidth]{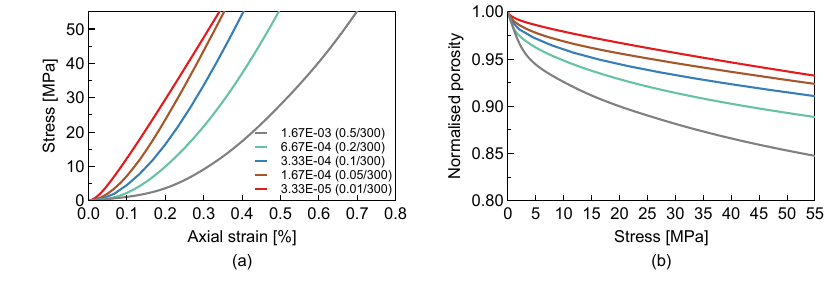}
\caption{Simulated stress-strain relationships and porosities resulted from applying different incremental strain conditions.
Different strains are achieved by applying different displacements (0.01, 0.05, 0.1, 0.2 and 0.5 pixel) divided by the entire domain length (300 pixels).}
\label{Senstivity}
\end{figure}

Figure \ref{Senstivity} illustrates the sensitivity analysis of the magnitude of applied macro strains during the deformation simulation.
Larger macro strain values lead to greater deformation, causing many partial-volume voxels to over-deform (pore fractions within these voxels have reduced below zero but cannot be converted to solid voxels in time due to the high macro strain increment). 
This results in a reduced slope in the stress-strain relationship and a lower normalised porosity. 
Conversely, as the applied macro strain is reduced, differences in the modelled results gradually decrease. 
Therefore, provided the macro strain increment is kept relatively low, the modelled results are expected to remain reliable.

%-----------------------------------------------------------
%-----------------------------------------------------------

%-----------------------------------------------------------
%-----------------------------------------------------------

\subsection{Uncertainties and recommendations for future work}
Like many other pore-scale modelling methods, the accuracy of the proposed method for rock deformation simulation is influenced by several factors. 
Firstly, image artefacts can affect the extraction of the partial-volume phase and the Beta distribution segmentation method. 
Although various filters and techniques have been developed to mitigate these artefacts, they may not always be effective, and their use introduces user bias and uncertainty \citep{Schluter2014, Andra2013b}. 
Additionally, the resolution of the images significantly impacts the modelling results. 
While treating voxels as partial-volume voxels with local pore fraction values can somewhat capture the closure of sub-resolution pore structures, if the image resolution is too low, the distinction between voxels containing sub-resolution pore spaces and pure solid voxels becomes less clear \citep{Li2024}. 
Consequently, these sub-resolution pore spaces may become indistinguishable and entirely lost in the reconstructed model.
Moreover, the use of effective medium theory models introduces a high degree of uncertainty. 
Different models often yield varying results, making the choice of an appropriate model crucial for more adequate simulation. 
Future research should focus on identifying more suitable models to enhance the accuracy of the simulation outcomes.
In addition, the presence of clay minerals in this work is neglected for simplicity. 
%This assumption may not significantly impact rocks with a low occurrence of clay minerals.
But the minor occurrence of clay minerals can significantly affect the effective elastic properties of the rock \citep{Han1986, Liang2020a}, and thereby the rock deformation performance.
Meanwhile, the presence of multiple minerals introduces additional complexity, making the identification of partial-volume phases significantly more challenging and affecting the accuracy of our simulation results. 
Therefore, incorporating multiple mineral phases into deformation modelling is an important consideration for future rock deformation simulations.
%However, it can lead to considerable discrepancies in the modelled results when applied to rocks with a high clay content.

One natural and important extension of this work would be to simulate permeability and analyse the pore-size distribution under varying stress conditions. 
However, the deformed rock model produced in this study contains voxels with a wide range of pore fractions. 
As a result, conventional methods for permeability calculations, such as the Lattice-Boltzmann method, and for pore-size distribution analysis, such as the maximum-ball algorithm, cannot be directly applied to these models, as they require binarised input data. 
A potential solution to this limitation could incorporate a down-scaling technique that converts voxels with varying pore fractions into binarised voxels at a higher image resolution. 
This would enable the use of conventional methods to simulate permeability and analyse pore-size distributions. 
In addition, using micro-continuum theory is also promising for computing the rock permeability \citep{Soulaine2016, Soulaine2024}.

\section{Conclusions}
This work proposes a novel method for simulating non-linear rock deformation under low-stress conditions. 
The approach begins with using the Beta distribution method to divide rock images into sub-phases with varying pore fractions. 
Next, effective medium theory and local pore fractions are applied to assign bulk and shear moduli to each sub-phase. 
The finite element method is then employed with incremental macro strains to simulate rock deformation, based on the assumption that the softer phase within a voxel deforms before the stiffer phase.

By calculating the porosity after deformation and regularising the deformed grids, this method effectively captures the non-linear behaviour of porosity changes, stress-strain relations, and rock elastic properties under low-stress levels. 
Visualising the rock models under different effective stresses reveals that many pore structures close under low stress. 
Larger pores shrink, and previously connected pores become less connected as stress increases. 
This pore closure effect makes the rock model stiffer, which leads to the non-linear deformation behaviour observed in porosity, stress-strain relationships, and rock elastic properties.

A natural extension of the current work is to integrate the proposed method with other techniques, such as down-scaling methods and micro-continuum theory. 
This would facilitate fluid transport simulations and allow for the analysis of pore space statistics in rocks under varying effective stress conditions.

\section*{Declaration of Competing Interest}
The authors declare that they have no known competing financial interests or personal relationships that could have appeared to influence the work reported in this paper.

\section*{Acknowledgement}
The financial support for this study was provided by PetroChina.

%\appendix
%\section{Example Appendix Section}
%\label{app1}

%%       \citep{<label>} ==> Lamport (1994)
%%       \citet{<label>} ==> (Lamport, 1994)
%%
%Example citation, See \citep{lamport94}.


  \bibliographystyle{elsarticle-harv} 
  \bibliography{Reference}



%\begin{thebibliography}{00}
%
%%% For authoryear reference style
%%% \bibitem[Author(year)]{label}
%%% Text of bibliographic item
%
%\bibitem[Lamport(1994)]{lamport94}
%  \textit{\LaTeX: a document preparation system},
%  Addison Wesley, Massachusetts,
%\end{thebibliography}

\end{document}